\documentclass[10pt,letterpaper]{article}
\usepackage{opex3}
\usepackage{mathptmx}
\usepackage{amsmath}
\usepackage{subfigure}
\usepackage{cite}
\usepackage[english]{babel}
\usepackage[T1]{fontenc}
\usepackage{graphicx}
\usepackage{amssymb}
\newcommand{\tfr}{FR_{z}}

\begin{document}
\renewcommand {\figurename}{Fig.}
\title{Self-healing in scaled propagation invariant beams}

\author{Victor Arriz\'on, Dilia Aguirre-Olivas, Gabriel Mellado-Villase\~nor and Sabino Ch\'avez-Cerda}

\address{Instituto Nacional de Astrof\'isica, \'Optica y Electr\'onica.\\Luis Enrique Erro \#1, 72840 Tonantzintla, Puebla, M\'exico}

\email{$^*$arrizon@inaoep.mx} 



\begin{abstract}
We analyze and demonstrate, numerically and experimentally,  the self-healing effect in scaled propagation invariant beams, subject to opaque obstructions. We introduce the signal to noise intensity ratio, a semi-analytical figure of merit, explicitly dependent on the features of the beams and the obstructions applied to them. The effect is quantitatively evaluated employing the Root Mean Square deviation and the similarity function.
\end{abstract}

\section{Introduction}
An optical beam that is subject to a partial obstruction propagates showing certain degree of self-reconstruction in the obstruction domain. This phenomenon is usually referred to as self-healing (SH) of the beam. The SH effect has been demonstrated and studied mainly in propagation invariant beams (PIBs), as Airy\cite{chris2008}, Bessel\cite{litvin,sabino,chu,bouchal}, Caustic\cite{sabino1}, Mathieu and Weber\cite{zhang} beams.

There are other interesting beams whose transverse intensity profile is also invariant under propagation, changing only their scale. We refer to these beams as scaled propagation invariant beams (SPIBs), to distinguish them from the PIBs. Examples of SPIBs are the Hermite-Gauss (HG), Ince-Gauss (IG), and Laguerre-Gauss (LG) beams, which are solutions of the paraxial scalar wave equation in three different coordinate systems\cite{verdeyen,bandres2004}. Considering that these beams show a kind of propagation invariance, it is also interesting to investigate their self-reconstruction capabilities. In this paper we analyze, demonstrate and evaluate the SH effect in SPIBs.

In section 2 we first recall the structure of SPIBs. Then we propose a method to analyze theoretically the SH effect in a generic SPIB, introducing the signal to noise intensity ratio, a semi-analytical figure of merit of the self-healing process, explicitly dependent on the features of the beam and the obstruction applied to it. This figure of merit, which is computed in base of analytical  diffraction formulas,  provides an indirect theoretical measurement of the degree of self-healing that can be expected for an obstructed beam. In addition to this indirect theoretical figure of merit, there is necessity of computing the true degree of self-reconstruction of propagated beams. This quantitative assessment is performed by means of the Root Mean Square (RMS) deviation and the similarity function (S) \cite{chu}. In section 3 we demonstrate and evaluate by means of numerical simulations and experimentally the SH in HG and IG beams.
\section{Theory}
\subsection{Structure of scaled propagation invariant beams}
The complex amplitude of the SPIBs (HG, IG and LG beams) includes a common factor with the modulation of a Gaussian beam, which is characterized by the Rayleigh length $z_{0}$. Other parameters, expressed in terms of $z_{0}$, are $w_{0}=(\lambda z_{0}/\pi)^{1/2}$, the beam radius waist, $w (z)=w_{0}[1+(z/z_{0})^{2}]^{1/2}$, the beam radius as a function of $z$, and $R(z)=z[1+(z_{0}/z)^{2}]$, the curvature radius of the quadratic phase.
The Gouy phase, which is proportional to $\arctan (z/z_{0})$, will be specified for each one of the beams. Next we describe the structure of SPIBs. For brevity we only present the analytical expressions for the HG and the IG beams.

The HG beam is the solution of the paraxial Helmholtz equation expressed in rectangular coordinates. Its complex amplitude is given by\cite{verdeyen}
\begin{equation}\label{eq:hgb}
HGB(x,y,z)=E_{0}\;\tfrac{w_{0}}{w(z)}\;H_{l}\left(\tfrac{\sqrt{2}}{w(z)} x\right)H_{n}\left(\tfrac{\sqrt{2}}{w(z)} y\right)\\
\exp\left[-\dfrac{r^{2}}{w(z)}-i k z +i\phi(z)-ik\dfrac{r^{2}}{2 R(z)}\right]
\end{equation}
where $H_{j}(j=l,n)$ is the $j$-th order Hermite polynomial, $r=\sqrt{x^{2}+y^{2}}$ is the radial coordinate, and $\phi(z)=(l+n+1)\arctan (z/z_{0})$ is the Gouy phase. The complex transmittance of the HG beam, at the waist plane ($z=0$) is expressed as
\begin{equation}\label{eq:hgb0}
HGB(x,y,z=0)=E_0\,H_{l}\left(\tfrac{\sqrt{2}}{w_{0}} x\right)H_{n}\left(\tfrac{\sqrt{2}}{w_{0}} y\right)\exp\left(-\tfrac{r^{2}}{w_{0}^{2}}\right)\,.
\end{equation}

The other SPIB to be considered, the IG beam, is the solution of the paraxial Helmholtz equation expressed in elliptical cylindrical coordinates. Such coordinates $(\xi,\eta,z)$ are defined by the relations $x=f(z)\cosh(\xi)\cos(\eta)$, $y=f(z)\sinh(\xi)\sin(\eta)$, and $z=z$, where $f(z)=f_{0}w(z)/w_{0}$ is the semifocal separation dependent on $z$ and $f_{0}$ is the semifocal separation at the waist plane. From these formulas it is established that the domains of $\xi$ and $\eta$ are $[0,\infty)$ and $[0,2\pi)$, respectively \cite{arfken}. The complex amplitudes of the even ($e$) and odd ($o$) IG beams are given by
\begin{equation}\label{eq:igbe}
IGB_{p,m}^{e}(\xi,\eta,z)=c\;\tfrac{w_{0}}{w(z)}\;C_{p}^{m}(i\xi,\epsilon)C_{p}^{m}(\eta,\epsilon)\\
\exp\left[-\dfrac{r^{2}}{w(z)}-i k z +i\phi(z)-ik\dfrac{r^{2}}{2 R(z)}\right],
\end{equation}
\begin{equation}\label{eq:igbo}
IGB_{p,m}^{o}(\xi,\eta,z)=s\;\tfrac{w_{0}}{w(z)}\;S_{p}^{m}(i\xi,\epsilon)S_{p}^{m}(\eta,\epsilon)\\
\exp\left[-\dfrac{r^{2}}{w(z)}-i k z +i\phi(z)-ik\dfrac{r^{2}}{2 R(z)}\right],
\end{equation}
where $C_{p}^{m}$ and $S_{p}^{m}$ are the Ince polynomials \cite{arscott} of order $p$ and degree $m$. The integer indices ($m,p$) that are in the ranges $0\leq m\leq p$ and $1\leq m\leq p$ for even and odd functions, respectively, must have the same parity, i. e. $(-1)^{p-m}=1$. The ellipticity parameter of IG beams is $\epsilon=2f_{0}^{2}/w_{0}^{2}$, and the Gouy phase is $\phi(z)=(p+1)\arctan (z/z_{0})$ \cite{bandres2004}. At the waist of the IG beams, their complex amplitudes reduce to
\begin{equation}\label{eq:igbe0}
    IGB_{p,m}^{e}(\xi,\eta,z=0)=c\;C_{p}^{m}(i\xi,\epsilon)C_{p}^{m}(\eta,\epsilon)
    \exp\left(-\tfrac{r^{2}}{w_{0}^{2}}\right),
\end{equation}
\begin{equation}\label{eq:igbo0}
    IGB_{p,m}^{o}(\xi,\eta,z=0)=s\;S_{p}^{m}(i\xi,\epsilon)S_{p}^{m}(\eta,\epsilon)
    \exp\left(-\tfrac{r^{2}}{w_{0}^{2}}\right),
\end{equation}
In Eqs. (\ref{eq:hgb}-\ref{eq:igbo0}), $E_{0}$, $c$ and $s$ are normalization factors.
\subsection{Analysis of self-healing in scaled propagation invariant beams}\label{sec:SHanalysis}
\begin{figure}[h!]
\centering\includegraphics[width=12cm]{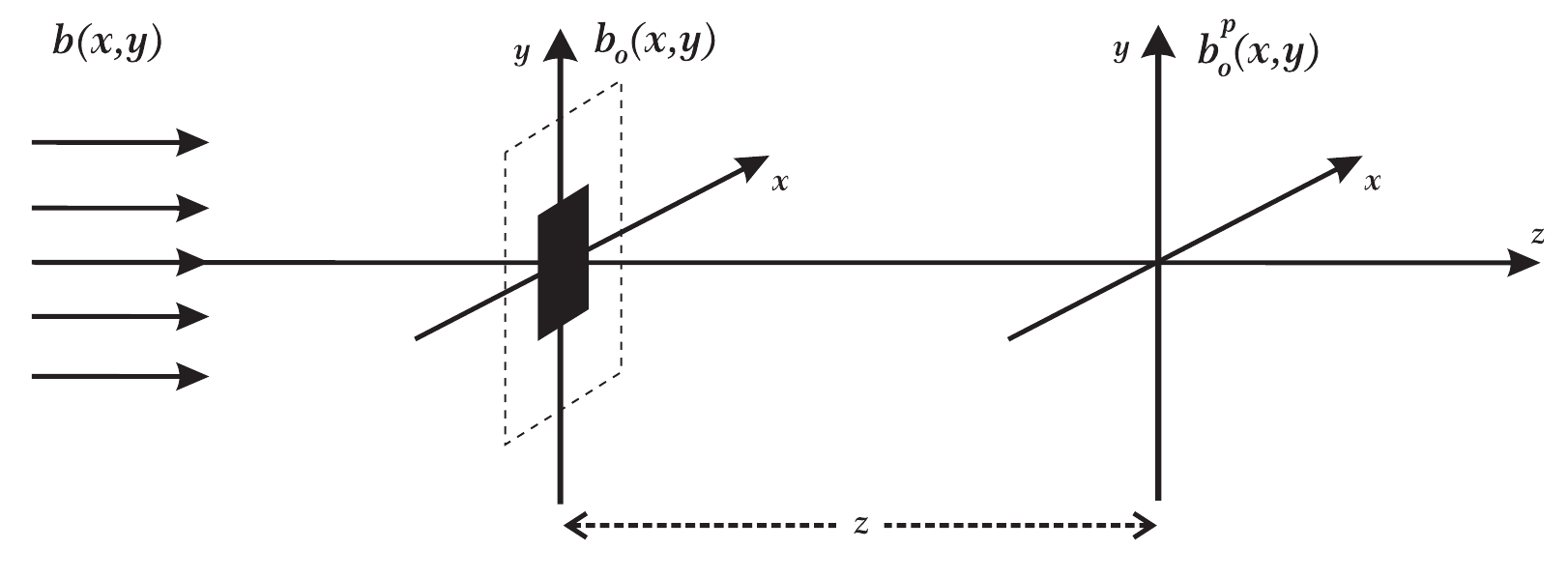}
	\caption{Optical setup: a beam of complex amplitude $b(x,y)$ partially obstructed at the plane $z=0$ is transformed into $b_{o}(x,y)$. Then, $b_{o}(x,y)$ propagates to the distance $z$ becoming $b_{o}^{p}(x,y)$.}
	\label{fig:setup1}
\end{figure}

To analyze the SH effect in SPIBs let us consider the setup depicted in Fig. \ref{fig:setup1}. It is assumed that a SPIB of complex amplitude $b(x,y)$, which arrives from the left side of the setup to the plane $z=0$, is partially obstructed at this plane.

The complex amplitude of the obstructed field at the plane $z=0$ can be expressed as
\begin{equation}\label{eq:bo}
    b_{o}(x,y)=b(x,y)[1-o(x,y)],
\end{equation}
where $o(x,y)$ is a binary function, equal to 1 at the obstruction area, and 0 otherwise. Let us denote as $b^{p}(x,y)$ the field that would propagate to the distance $z$ if the beam $b(x,y)$ were not obstructed. This field can be expressed as
\begin{equation}\label{eq:bp}
    b^{p}(x,y)=\tfr\{b(x,y)\},
\end{equation}
where $\tfr$ is the operator that represents Fresnel free propagation to a distance $z$. The field that propagates to the distance $z$ when the obstruction is present, can be expressed as
\begin{equation}\label{eq:bop}
    b_{o}^{p}(x,y)=\tfr\{b_{o}(x,y)\}=b^{p}(x,y)-n(x,y),
\end{equation}
where
\begin{equation}\label{eq:noise}
    n(x,y)=\tfr\{b(x,y)o(x,y)\}.
\end{equation}

According to Eq. \eqref{eq:bop}, the non-obstructed propagated field  $b^{p}(x,y)$ appears as a part of the obstructed propagated field. The other function in Eq. \eqref{eq:bop}, $n(x,y)$, plays the role of a perturbation of the non-obstructed field. To analyze the SH effect it will be helpful to know the structure of $n(x,y)$. This can be done applying in Eq. (\ref{eq:noise}) the formula for the Fresnel transform of a product of functions \cite{arrizon2001}, to obtain
\begin{equation}\label{eq:noise2}
n(x,y)=\dfrac{E(x,y,z)}{\lambda^{2}z^{2}}\{E^{*}(x,y,z)b^{p}(x,y)\}\otimes O \left(\tfrac{x}{\lambda z},\tfrac{y}{\lambda z}\right),
\end{equation}
where $O(u,v)$ is the Fourier transform of the obstruction pupil $o(x,y)$, $\otimes$ denotes the convolution operation, and $E(x,y)=\exp[(i\pi/\lambda z)(x^2+y^2)]$. According to Eq. \eqref{eq:bop}, a good approximation of the non-obstructed field $b^p(x,y)$ will be recovered in the propagated field $b_{o}^p(x,y)$ if the intensity of $n(x,y)$ is negligible in comparison to the intensity of $b^p(x,y)$. Next we establish estimated values of such intensities and propose a criterion for self-reconstruction and the propagation distance required to fulfill it. To determine the intensities we require optical powers  of the involved fields [in Eq. (\ref{eq:bop})] and the areas where these fields are distributed. We first establish the transverse widths of the field  $b^p(x,y)$ as
\begin{align}\label{eq:widths}
\Delta x_b&=2\alpha_xw(z)\notag\,,\\
\Delta y_b&=2\alpha_yw(z)\,,
\end{align}
where $\alpha_x$ and $\alpha_y$ are constant parameters and $w(z)$ is the radius of the beam Gaussian factor, at the propagation distance $z$. To establish the widths in Eq. \eqref{eq:widths}, and the corresponding parameters $\alpha_x$ and $\alpha_y$, we specify the beam limits (along any horizontal or vertical line) at the positions where the edge intensities reduces by a factor of $1/e^2$ (respect to the peak intensity). The horizontal field
width is the maximum of the widths obtained for the different horizontal lines. A similar criterion provides the vertical field  width. It is also important to establish the transverse widths $\Delta x_o$ and $\Delta y_o$, of the function $O(x/\lambda z,y/\lambda z)$ that appears in Eq. (\ref{eq:noise2}). As particular case we assume that $o(x,y)$ is a square obstruction, of width $a$, for which we obtain
\begin{equation}
\Delta x_o = \Delta y_o = {2\lambda z}/{a}\,.
\label{eq:widthSinc}
\end{equation}

Considering general features of the convolution operation in Eq. (\ref{eq:noise2}) we estimate that the transverse widths of the perturbation field $n(x,y)$, along the horizontal and vertical axes, are respectively
\begin{align}\label{eq:extention}
\Delta x_n&=\Delta x_b+\Delta x_o \notag\,,\\
\Delta y_n&=\Delta y_b+\Delta y_o\,.
\end{align}

The areas of the domains of $b^p(x,y)$ and $n(x,y)$, in terms of the widths defined in Eqs. \eqref{eq:widths} and \eqref{eq:extention},  are given by
\begin{align}\label{eq:areas}
A_b&=\Delta x_b\Delta y_b\notag \,,\\
A_n&=\Delta x_n\Delta y_n\,.
\end{align}
On the other hand, we denote the optical powers of $b^p(x,y)$ and $n(x,y)$ as $P_b$ and $P_n$. Considering that such powers are invariant during propagation, we can compute them at the plane $z=0$. The powers $P_b$ and $P_n$ results from integrating $\left|b(x,y)\right|^2$ in the whole plane $z=0$ and in the obstruction domain respectively. Therefore, we can establish the average intensities $\overline{I}_{b}=P_b/A_b$ and $\overline{I}_{n}=P_n/A_n$, which correspond to the field $b^p(x,y)$ and the perturbation function $n(x,y)$, respectively. Now we can introduce the parameter

\begin{equation}\label{eq:cSH}
Q\equiv\frac{\overline{I}_{b}}{\overline{I}_{n}}\,,
\end{equation}
a positive quantity, which is referred to as signal to noise intensity ratio. An important result, proved below, is that the ratio $Q$ is limited by an upper bound, explicitly dependent on the parameters of the beam and the obstruction applied to it. Introducing the relative power $P_{nb}=P_n/P_b$, Eq. (\ref{eq:cSH}) can be expressed as
{\begin{equation}\label{eq:cSH1}
\frac{A_n}{A_b}=Q P_{nb}\,.
\end{equation}

Considering Eqs. \eqref{eq:widths} to \eqref{eq:areas} and performing some algebra, Eq. (\ref{eq:cSH1}) leads to the relation
\begin{equation}
\frac{\Delta x_o}{\Delta x_b}=\sqrt{Q\alpha P_{nb}+\alpha_d^2}-\alpha_p\,,
\label{eq:DxoDxb1}
\end{equation}
where $\alpha=\alpha_y/\alpha_x$, $\alpha_p=(\alpha+1)/2$ and $\alpha_d=(\alpha-1)/2$. Considering that ${\Delta x_o}/{\Delta x_b}>0$, we obtain the inequality
\begin{equation}\label{eq:minimumB}
\frac{\alpha_p^2-\alpha_d^2}{\alpha P_{nb}}<Q\,,
\end{equation}
that provides a minimum bound for $Q$. On the other hand, considering definitions for $\Delta x_o$ and $\Delta x_b$, Eq. \eqref{eq:DxoDxb1} can be transformed into
{\begin{equation}
\frac{z}{z_o}=\left[\frac{\pi^2w_o^2\left(\sqrt{Q\alpha P_{nb}+\alpha_d^2}-\alpha_p\right)^{-2}}{a^2\alpha_x^2}-1\right]^{-1/2}\,,
\label{eq:zzo}
\end{equation}
which corresponds to the propagation distance required to obtain a given signal to noise intensity ratio $Q$. Now, since the expression under the square brackets in Eq. \eqref{eq:zzo} must be positive, one obtains the inequality
\begin{equation}\label{eq:Q}
Q<\frac{1}{\alpha P_{nb}}\left[\left(\alpha_p+\frac{\pi w_o}{a\alpha_x}\right)^2-\alpha_d^2\right]\,,
\end{equation}
that gives a maximum bound for $Q$, which complements the minimum bound, given by \eqref{eq:minimumB}. It is noted that the maximum value of $Q$, in  \eqref{eq:Q}, is dependent on the obstruction size $a$, the relative power $P_{nb}$, and the beam features related to the parameters $\alpha$, $\alpha_x$, $\alpha_p$ and $\alpha_d$. As a consequence of this result, given the parameters that specify the beam and the obstruction applied to it, there is a limit for the parameter $Q$, which corresponds indirectly to a limit for the degree of SH that can be attained on propagation. When the beam is not obstructed both parameters $a$ and $P_{nb}$ are null and the upper bound for $Q$ becomes $\infty$. In any other case, the upper bound value of $Q$ is finite.
\section{Numerical and experimental results}
To evaluate the validity of the quantitative formulae proposed in the previous section we study the SH of HG and IG SPIBs. For the experimental generation of the beams under test we employ amplitude computer-generated holograms (CGHs), displayed in a twisted nematic liquid crystal (TNLC) spatial light modulator (SLM)\cite{arrizon2005}. The holograms are designed to include the obstruction as a feature of the generated fields, simplifying the experimental set-up. Let us briefly describe the features of these CGHs. We assume that the CGH is used to encode the optical field $s(x,y)=a(x,y)\exp(i\phi(x,y))$ where the amplitud $a(x,y)$ is a normalized positive function and the phase $\phi(x,y)$ takes values in the range $[-\pi,\pi]$. The transmittance of the amplitude CGH that allows the generation of the field $s(x,y)$ is given by
\begin{equation}\label{eq:cgh}
    h(x,y)=h_{b}(x,y)+a(x,y)\cos\left[\phi(x,y)-i2\pi(u_{0}x+v_{0}y)\right],
\end{equation}
where $(u_0,v_0)$ are the spatial frequencies of a linear phase carrier and $h_b(x,y)$ is a background function that makes positive definite the function $h(x,y)$ \cite{arrizon2005}. In order to propitiate a high signal to noise ratio in the generation of the field $s(x,y)$ we must chose a background $h_b(x,y)$ with low power and low  bandwidth. In the CGHs that we implement, $h_b(x,y)$ is chosen as a soft Gaussian function.

The Fourier spectrum of the CGH in Eq. \eqref{eq:cgh} is
\begin{equation}
    H(u,v)=H_b(u,v)+\dfrac{1}{2}S(u+u_{0},v+v_{0})+\dfrac{1}{2}S^{*}(-u-u_{0},-v-v_{0})
\end{equation}
where $H_b(u,v)$ is the Fourier transform of $h_b(x,y)$, $S$ is the Fourier transform of $s(x,y)$, and $S^{*}$ is its complex conjugate. We assume that $S(u,v)$ corresponds to one of the desired SPIBs (HG or IG beams).  To isolate the SPIB, a band-pass filter, centered at frequency coordinates $(-u_0,-v_0)$, is placed at the Fourier domain of the CGH.

The experimental setup designed to generate the SPIBs, is depicted in Fig. \ref{fig:setup}. In this setup, an expanded and collimated He-Ne laser beam is used to illuminate the CGHs codified on a TNLC-SLM. Linear polarizers $P_{1}$ and $P_{2}$, orthogonal to each other, are required to obtain mostly amplitude modulation in the SLM. The Fourier transform of the amplitude CGHs  is generated by the lens $L_{3}$. The field $S(u+u_0,v+v_0)$, consisting in one of the SPIBs (HG or IG beams) appears at the open pupil in the spatial filter (SF) plane. For convenience, a dark area corresponding to the required obstruction, is optionally encoded in the SPIB domain. The intensities of the generated fields are recorded with a CCD camera at different distances along the propagation axis.
\begin{figure}[h!]
\centering\includegraphics[width=14cm]{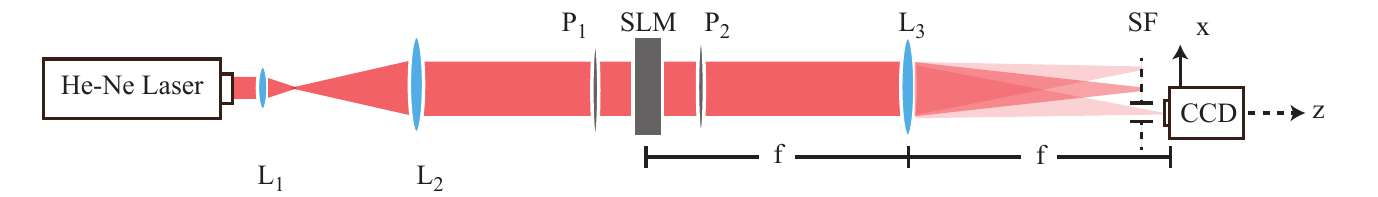}
	\caption{Experimental setup to demostrate self-healing of HG and IG beams. These fields, that include optionally a dark area (or obstruction area), appear at the first order in the Fourier domain of the CGH encoded in the TNLC-SLM. The CGH Fourier transform is generated by the lens $L_{3}$. The fields are recorded with a CCD camera at different distances $z$. }
	\label{fig:setup}
\end{figure}
\newline

In Fig. \ref{fig:sinobs} we show the intensities generated numerically of a HG beam (a) and an odd IG beam (b) that will be employed to illustrate the SH effect in SPIBs. The parameters for the HG beam are $l=n=8$, and for the odd IG beam are $p=8$ and $m=2$. The intensities of the beams,  experimentally generated with amplitude CGHs, using the optical setup in Fig. \ref{fig:setup}, are shown in Fig. 3(c,d). The focal length of the lens $L_{3}$ was $f=75 cm$, and the waists of the generated beams, at the plane of the SF, were $200 \mu$ and $137 \mu$ respectively.

\begin{figure}[h!]
\centering
        \subfigure[]{
          \includegraphics[width=2.2cm]{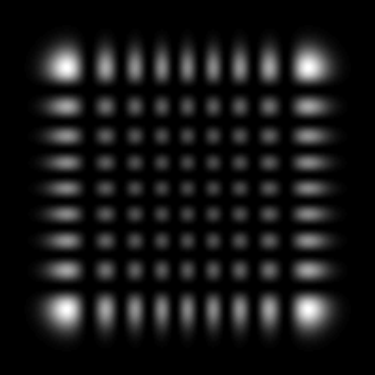}
          \label{fig:Nhgb88}}
        \subfigure[]{
          \includegraphics[width=2.2cm]{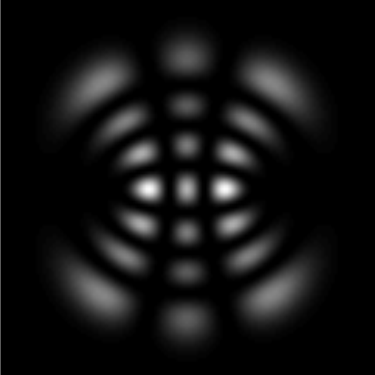}
          \label{fig:Nigb823}}
        \subfigure[]{
          \includegraphics[width=2.2cm]{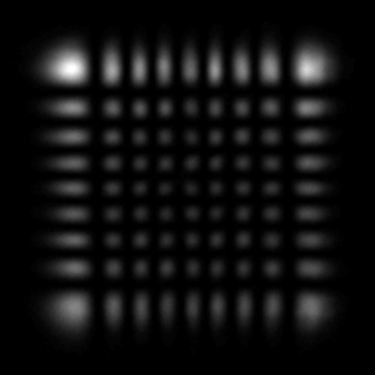}
          \label{fig:hgb88}}
        \subfigure[]{
          \includegraphics[width=2.2cm]{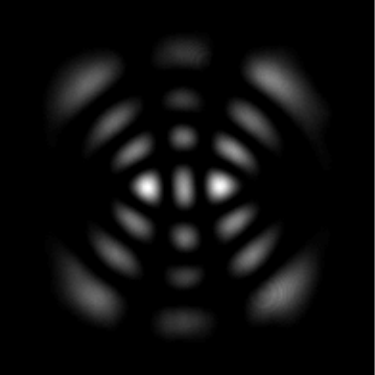}
          \label{fig:igb823}}
\caption{(a, b) Numerical and (c, d) experimental intensities of HG and IG beams, employed to illustrate the SH effect in SPIBs.}\label{fig:sinobs}
\end{figure}
In order to test the SH effect in partially obstructed SPIBs, the CGHs used to generate the fields in Fig. \ref{fig:sinobs},  were modified to include different dark square areas (obstruction domains), at the center of each field. The intensities of the obtained fields, at  different propagation distances, are displayed in figures \ref{fig:NH88om} to \ref{fig:Ni823op}. For each figure the results corresponding to the numerical simulations and the experiments appear at the  top and the bottom images respectively. For each horizontal array of images, in any of the figures,  the field at z=0 (at the left extreme of the array) presents the obstructed domain in focus. The widths of the fields of view, in the different images, are proportional to the width $w(z)$ of the Gaussian factor, at the corresponding distance $z$.

In Fig. \ref{fig:NH88om} we present the results for the HG beam, with indices $l=n=8$. The waist radius and obstacle width in this case are respectively $w_{o}=200\, \mu m$ and $a=2.25\;w_{o}$. Employing the definitions in section \ref{sec:SHanalysis}, we obtained the parameters $\alpha_{x}=\alpha_{y}=3$ and $P_{nb}\approx0.1$.  For these parameters we employed Eqs. \eqref{eq:minimumB} and \eqref{eq:Q} to obtain the  allowed values of $Q$ in the range [9.90:18.72). For each $Q$ in this allowed range we can compute the required distance $z/z_o$, using Eq. \eqref{eq:Q}.  For example, if we chose the ratio $Q = 16.5$, the computed normalized propagation distance obtained is $z/z_{o}=1.2$. It is noticed that the upper value of $Q$ in the interval is avoided, since in this case $z$ diverges to infinity. Both, the numerical and experimental results in Fig. \eqref{fig:NH88om} correspond to normalized propagation distances ($z/z_{o}$) in the range $[0:1.2]$. The propagation distance for the fields in Fig. \ref{fig:NH88om} marked as (d) and (h) correspond to the signal to noise intensity ratio $Q=16.5$.
\begin{figure}[h!]
\centering
	\subfigure[]{
  	\includegraphics[width=2.2cm]{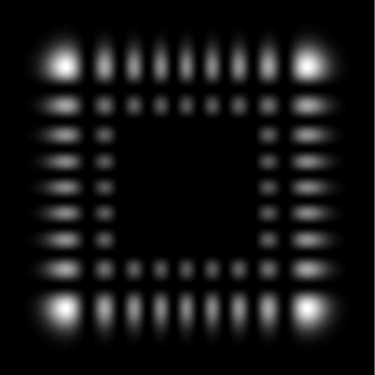}
  	\label{fig:NH88om0cm}}
 	\subfigure[]{
    \includegraphics[width=2.2cm]{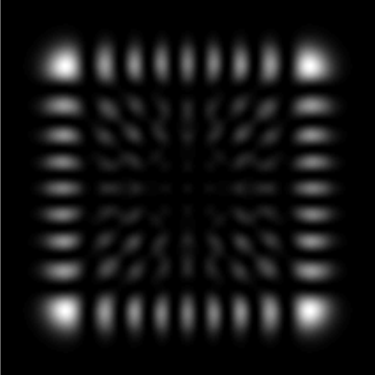}
   	\label{fig:NH88om8cm}}
 	\subfigure[]{
   	\includegraphics[width=2.2cm]{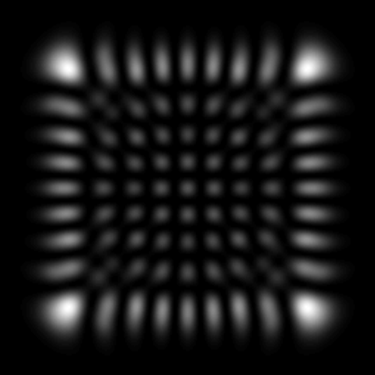}
   	\label{fig:NH88om16cm}}
 	\subfigure[]{
   	\includegraphics[width=2.2cm]{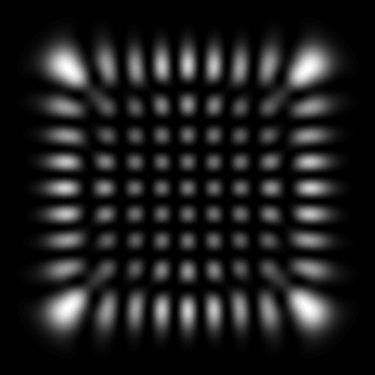}
   	\label{fig:NH88om24cm}}
   \hspace{3cm}
 	\subfigure[]{
   	\includegraphics[width=2.2cm]{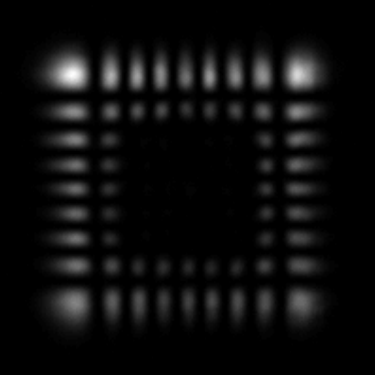}
   	\label{fig:00cmom}}   	
	\subfigure[]{
   	\includegraphics[width=2.2cm]{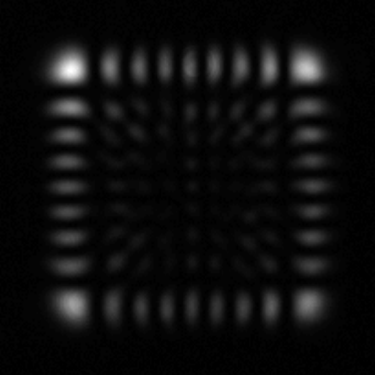}
   	\label{fig:08cmom}}
	\subfigure[]{
   	\includegraphics[width=2.2cm]{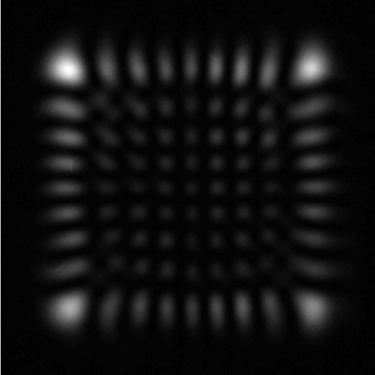}
   	\label{fig:16cmom}}
	\subfigure[]{
   	\includegraphics[width=2.2cm]{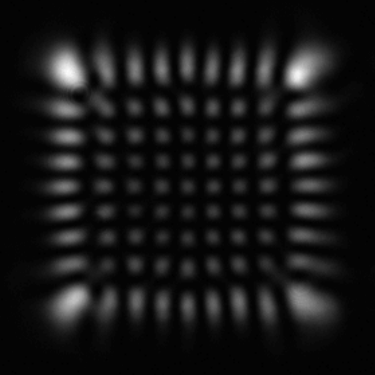}
   	\label{fig:24cmom}}
\caption{Numerical (top images) and experimental (bottom images) results for self-healing of a HG beam subject to a medium sized square obstruction, at the propagation distance $z/z_{0}$: (a,e) $0$, (b,f) $0.4$, (c,g) $0.8$, and (d,h) $1.2$.}
\label{fig:NH88om}
\end{figure}
The results obtained for the same HG beam employing an increased obstruction area of size $a=4.16\;w_{o}$, are displayed in Fig. \ref{fig:NH88og}. In this case, the parameters $\alpha_x$ and $\alpha_y$ remain unchanged and the relative powers take the increased value $P_{nb}=0.25$, obtaining the allowed values of $Q$ in the range $(4.03:6.32)$. The different recorded fields (numerical and experimental) correspond to propagation distances $z/z_{o}$ in the range $[0:1.6]$, with incremental step $0.4$. In particular the propagation distance $z/z_{o}=1.6$ corresponds to the signal to noise intensity ratio $Q=5.94$. This Q value, and the associated propagation distance, correspond to the fields displayed in the images marked as (e) and (j).

\begin{figure}[h!]
\centering
	\subfigure[]{
  	\includegraphics[width=2.2cm]{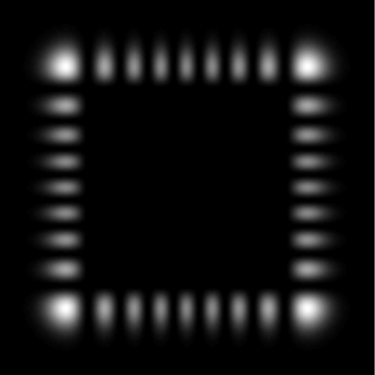}
  	\label{fig:NH88og0cm}}
 	\subfigure[]{
 	\includegraphics[width=2.2cm]{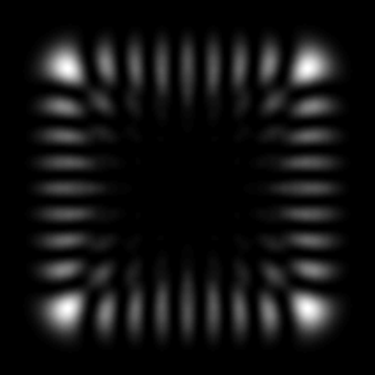}
   	\label{fig:NH88og8cm}}
 	\subfigure[]{
   	\includegraphics[width=2.2cm]{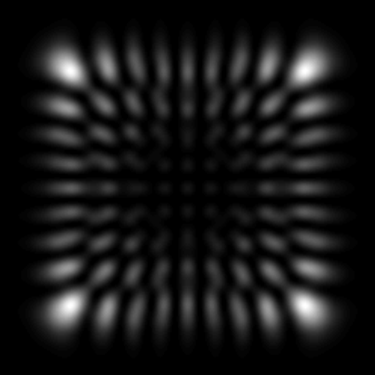}
   	\label{fig:NH88og16cm}}
 	\subfigure[]{
   	\includegraphics[width=2.2cm]{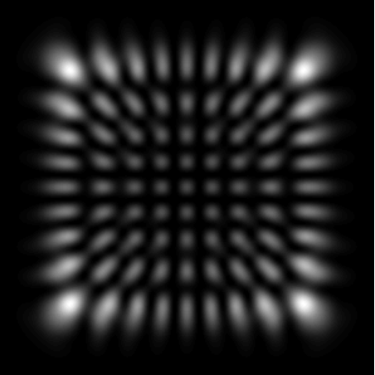}
   	\label{fig:NH88og24cm}}
 	\subfigure[]{
   	\includegraphics[width=2.2cm]{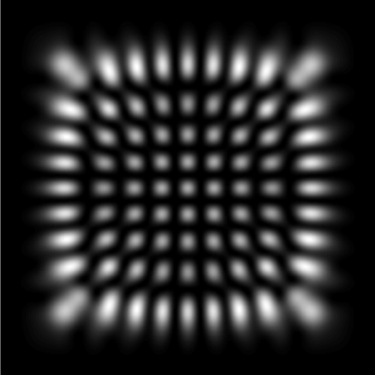}
   	\label{fig:NH88og32cm}}   	
    \hspace{3cm}
 	\subfigure[]{
   	\includegraphics[width=2.2cm]{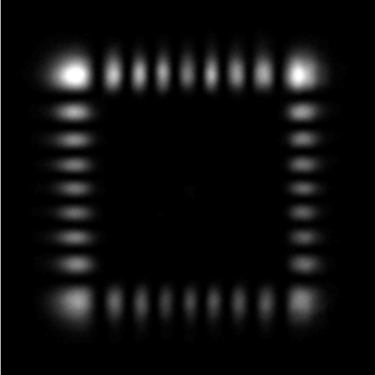}
   	\label{fig:00cmog}}
    \subfigure[]{
   	\includegraphics[width=2.2cm]{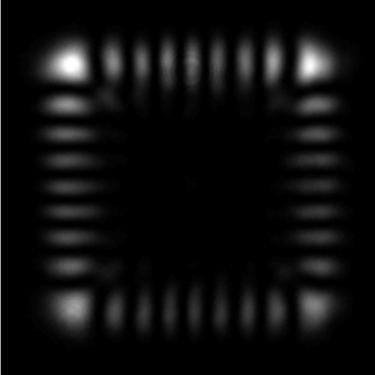}
   	\label{fig:08cmog}}
    \subfigure[]{
   	\includegraphics[width=2.2cm]{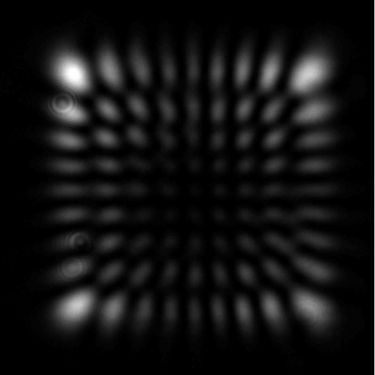}
   	\label{fig:18cmog}} 	
    \subfigure[]{
   	\includegraphics[width=2.2cm]{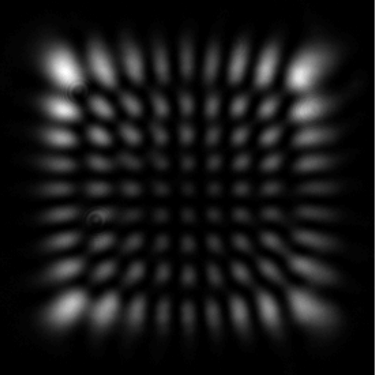}
   	\label{fig:24cmog}}
    \subfigure[]{
   	\includegraphics[width=2.2cm]{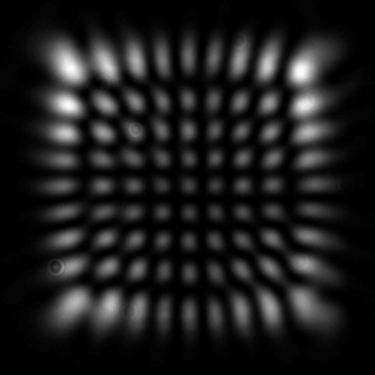}
   	\label{fig:32cmog}}   	
\caption{Numerical (top images) and experimental (bottom images) results for self-healing of a HG beam subject to a relative large square obstruction, at the propagation distance $z/z_{0}$: (a,f) $0$, (b,g) $0.4$, (c,h) $0.8$, (d,i) $1.2$, and (e,j) $1.6$.}
\label{fig:NH88og}
\end{figure}

In the case of the  IG beam,  the SH demonstration is performed with an obstruction of width $a=2.7 \;w_{o}$ (where $w_{o}=137 \; \mu m$). Others parameters of the IG beam are the constants $\alpha_{x}=2$, $\alpha_{y}=3$, and $P_{nb}=0.358$, for which we obtained the allowed values of $Q$ in the range $(2.79:7.02)$. The results (numerical end experimental) displayed in Fig. \ref{fig:Ni823op} correspond to propagation distances $z/z_{o}$ in the range $[0:1.28]$ with approximated steps of $0.43$. The propagation distance $z/z_{o}=1.28$, in this range, correspond to the intensity ratio $Q=5.34$ (smaller than the limit value $Q=6.16$). In the numerical and experimental results, displayed in figures \ref{fig:NH88om} to \ref{fig:Ni823op},  it is noticed a relatively good self-reconstruction of the field inside the obstruction domain. In contrast, the field outside the obstruction area has been clearly affected by the perturbation term $n(x,y)$. This qualitative observation is verified by the quantitative assessment of the propagated fields, in the next subsection.
\begin{figure}[h!]
\centering
	\subfigure[]{
  	\includegraphics[width=2.2cm]{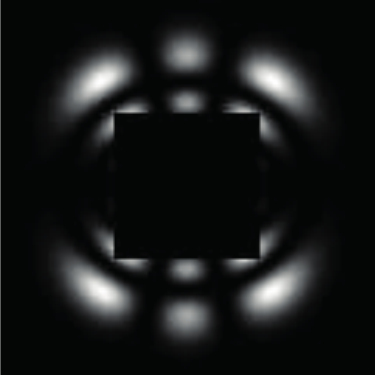}
  	\label{fig:igb823op0cm}}
 	\subfigure[]{
 	\includegraphics[width=2.2cm]{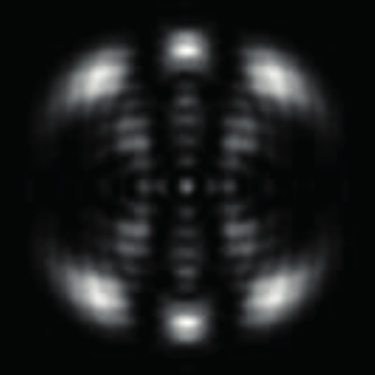}
   	\label{fig:igb823op4cm}}
 	\subfigure[]{
   	\includegraphics[width=2.2cm]{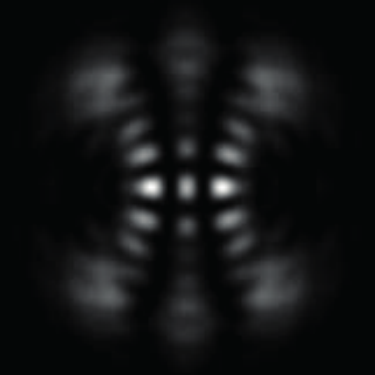}
   	\label{fig:igb823op8cm}}
 	\subfigure[]{
   	\includegraphics[width=2.2cm]{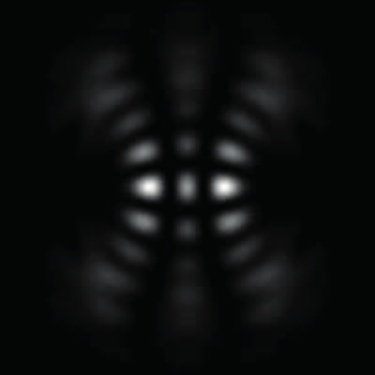}
   	\label{fig:igb823op12cm}}
    \hspace{4cm}
    \subfigure[]{
  	\includegraphics[width=2.2cm]{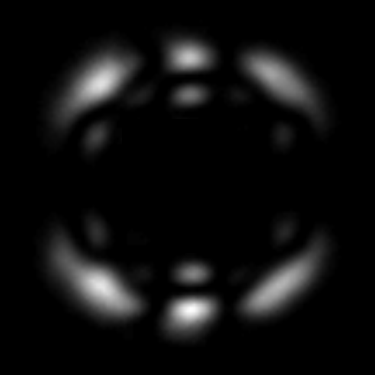}
  	\label{fig:igb823op0cm}}
 	\subfigure[]{
 	\includegraphics[width=2.2cm]{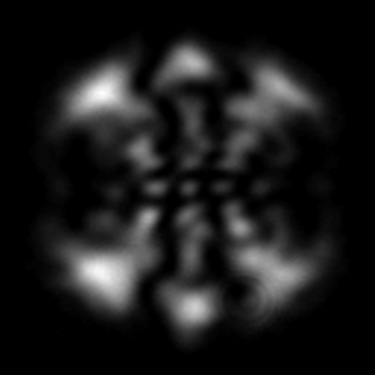}
   	\label{fig:igb823op4cm}}
 	\subfigure[]{
   	\includegraphics[width=2.2cm]{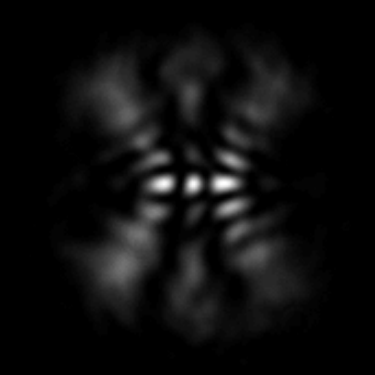}
   	\label{fig:igb823op8cm}}
 	\subfigure[]{
   	\includegraphics[width=2.2cm]{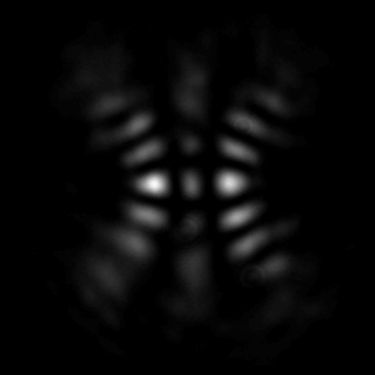}
   	\label{fig:igb823op12cm}}
\caption{Numerical (top images) and experimental (bottom images) results for self-healing of a IG beam subject to a small sized square obstruction, at the propagation distance $z/z_{0}$: (a,e) $0$, (b,f) $0.43$, (c,g) $0.85$, and (d,h) $1.28$.}
\label{fig:Ni823op}
\end{figure}
\subsection{Quantitative evaluation of self-healing}
In section \ref{sec:SHanalysis} we introduced the parameter $Q$ that represents a rough form of signal to noise ratio in the self-reconstruction of the desired field $b^p$, during propagation of the obstructed field $b_o^p$. As a complement of this semi-analytical assessment of the self-healing process, here we evaluate this process employing two figures of merit: the Root Mean Square (RMS) deviation and the Similarity (S) function\cite{chu}.

The RMS deviation of the intensity of the obstructed field, $I^{p}_{o}=|b^{p}_{o}|^2$, respect to intensity of the non-obstructed field, $I^{p}=|b^{p}|^2$, is given by:
\begin{equation}
    RMS=\frac{\int \limits _{w}\left(I^{p}-\beta I^{p}_{o}\right)^{2}\;dA}{\int \limits _{w} (I^{p})^{2}\;dA}
\end{equation}
where $dA$ is the differential area, $w$ is the domain where the RMS is evaluated, and $\beta$ is a constant that allows the best fitting of the intensities $I_o^p$ and $I^p$, obtained from the relation, $\partial RSM / \partial \beta=0$. On the other hand, the similarity is defined as:
 \begin{equation}
    S=\frac{(|b^{p}_{o}|,|b^{p}|)}{||b^{p}_{o}||\;\;||b^{p}||},
\end{equation}
where $(|b^{p}_{o}|,|b^{p}|)=\int _{w}(b^{p}_{o})\;(b^{p})^{*}dA$ is the inner product of two fields, meanwhile $||b^{p}_{o}||=[\int _{w}(b^{p}_{o})\;\;(b^{p}_{o})^{*}]^{1/2}dA$ and $||b^{p}||=[\int _{w}(b^{p})\;\;(b^{p})^{*}]^{1/2}dA$ are the norm of the fields, and asterisk ($^*$) denotes complex conjugation. For each propagation distance $(z)$, the RMS and S are computed in a square domain $\varOmega$ of size $a(z)=a(w(z)/w_{o})$, referred to as field internal domain. The scale change in this domain is equal to that of the beam itself, during its free propagation. Additionally, we evaluated numerically the $RMS$ and $S$ for a domain external to $\varOmega$.

Values for the RMS deviation and the similarity S in the domain $\varOmega$, for the HG beams in Fig. \ref{fig:Nhgb88}, subject to different centered square obstructions, and propagated to different distances, are  displayed in Fig. \ref{fig:rmssimai}(a,b). In addition to the medium and large obstruction, which correspond to the ones employed in the numerical and experimental results (for the HG beam) just discussed, we also considered here a small obstruction, of width $a=1.68\;w_{o}$. Similar results for the IG beam in Fig. \ref{fig:Nigb823}, employing square obstructions of two different widths, are displayed in Fig. \ref{fig:rmssimai}(c,d). In this case the medium and large obstructions have widths $2.80\;w_{o}$ and $4.16\;w_{o}$, respectively (with $w_{o}=200\mu m$).  As noted in the results of Fig. \ref{fig:rmssimai}, the behavior of the similarity is opposite to that of the RMS deviation. It is also noted that the RMS metric presents more sensible variations along the propagation ranges. An interesting fact is that the RMS attain its best (minimum) value at certain propagation distances. E.g. the RMS for the HG beam with the medium sized obstruction shows a minimum at $z/z_{o} \sim 2.3$. Another case of the IG beam, with the medium sized  obstruction that shows a minimum RMS at $z/z_{o}\sim 2.2$. This behavior is consistent  with the prediction in section \ref{sec:SHanalysis} of an upper bound value for the signal to intensity ratio $Q$.
\begin{figure}[h!]
\centering
	\subfigure[]{
  	\includegraphics[width=6.4cm]{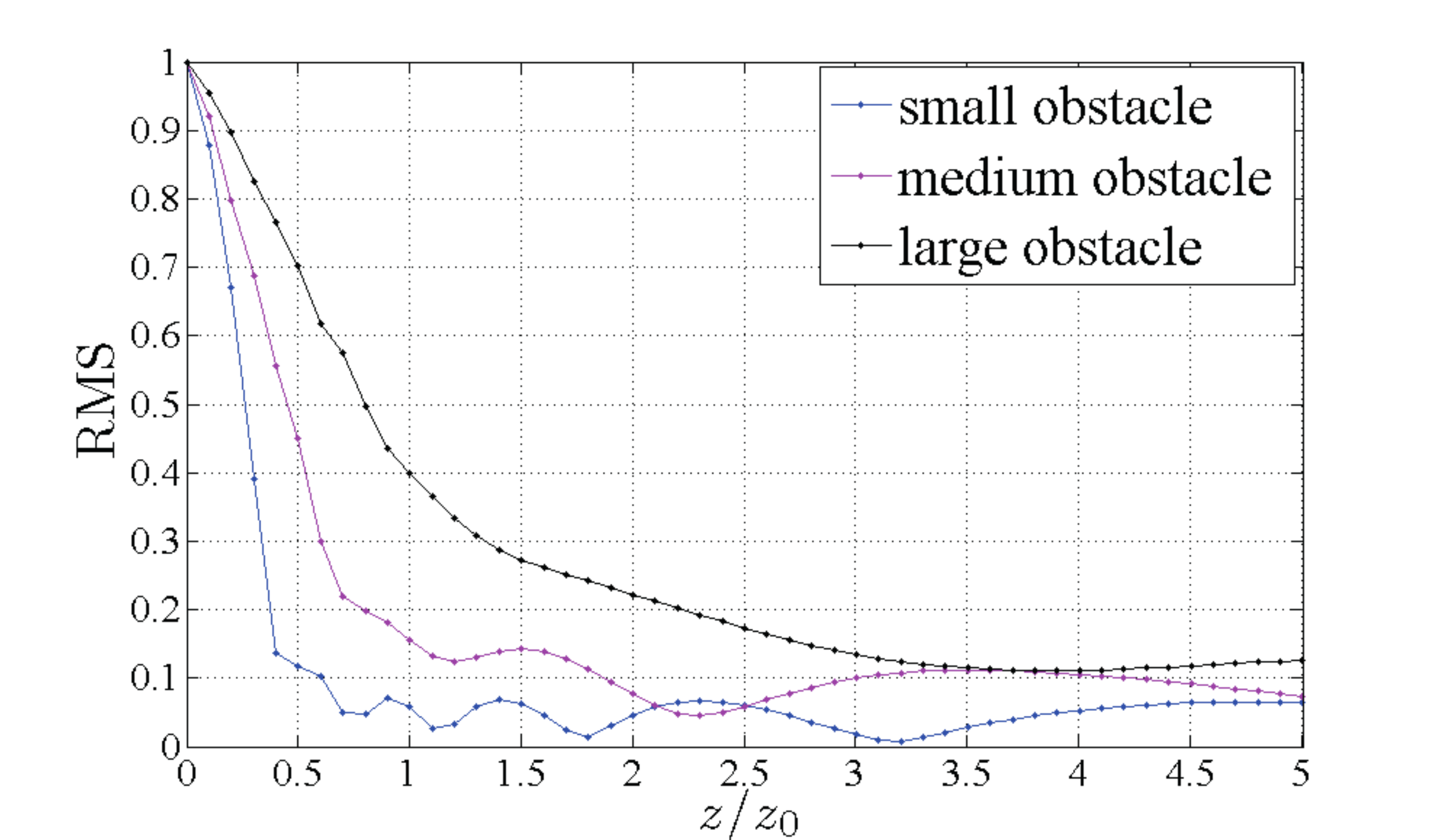}
  	\label{fig:rmshgb_ai}}
    \subfigure[]{
  	\includegraphics[width=6.4cm]{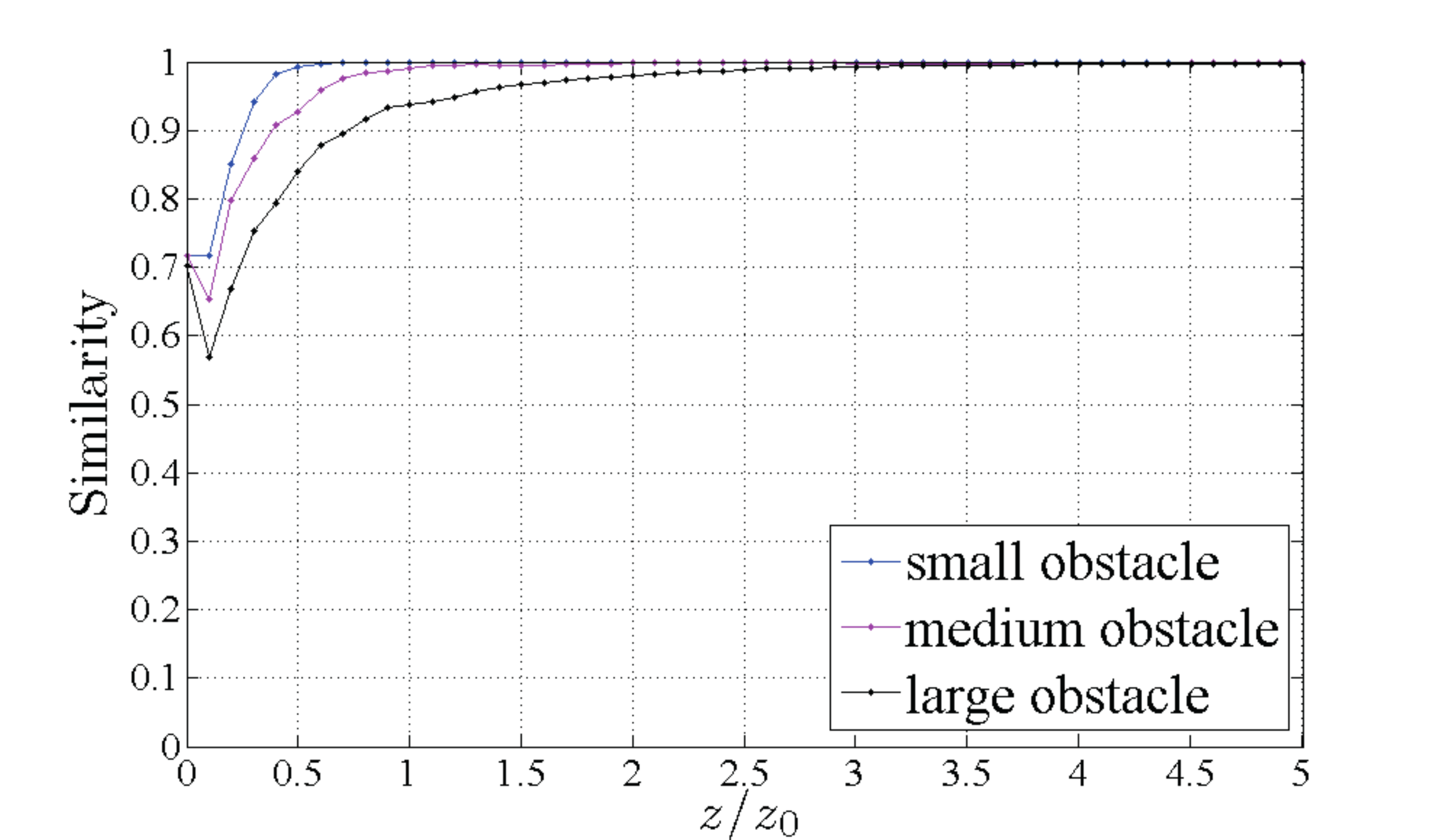}
  	\label{fig:simihgb_ai}}
 	\subfigure[]{
 	\includegraphics[width=6.4cm]{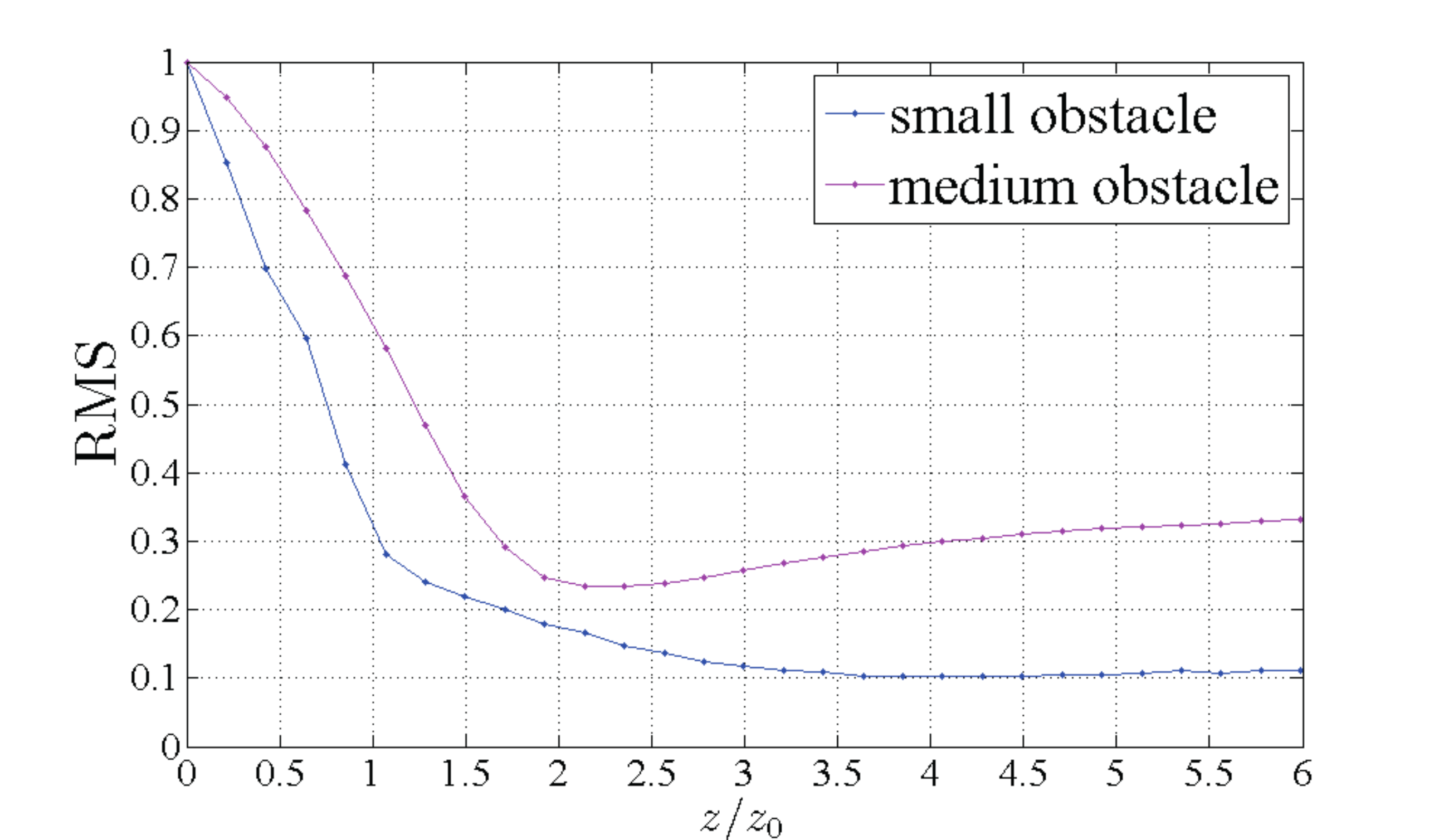}
   	\label{fig:rmsigb_ai}}
 	\subfigure[]{
  	\includegraphics[width=6.4cm]{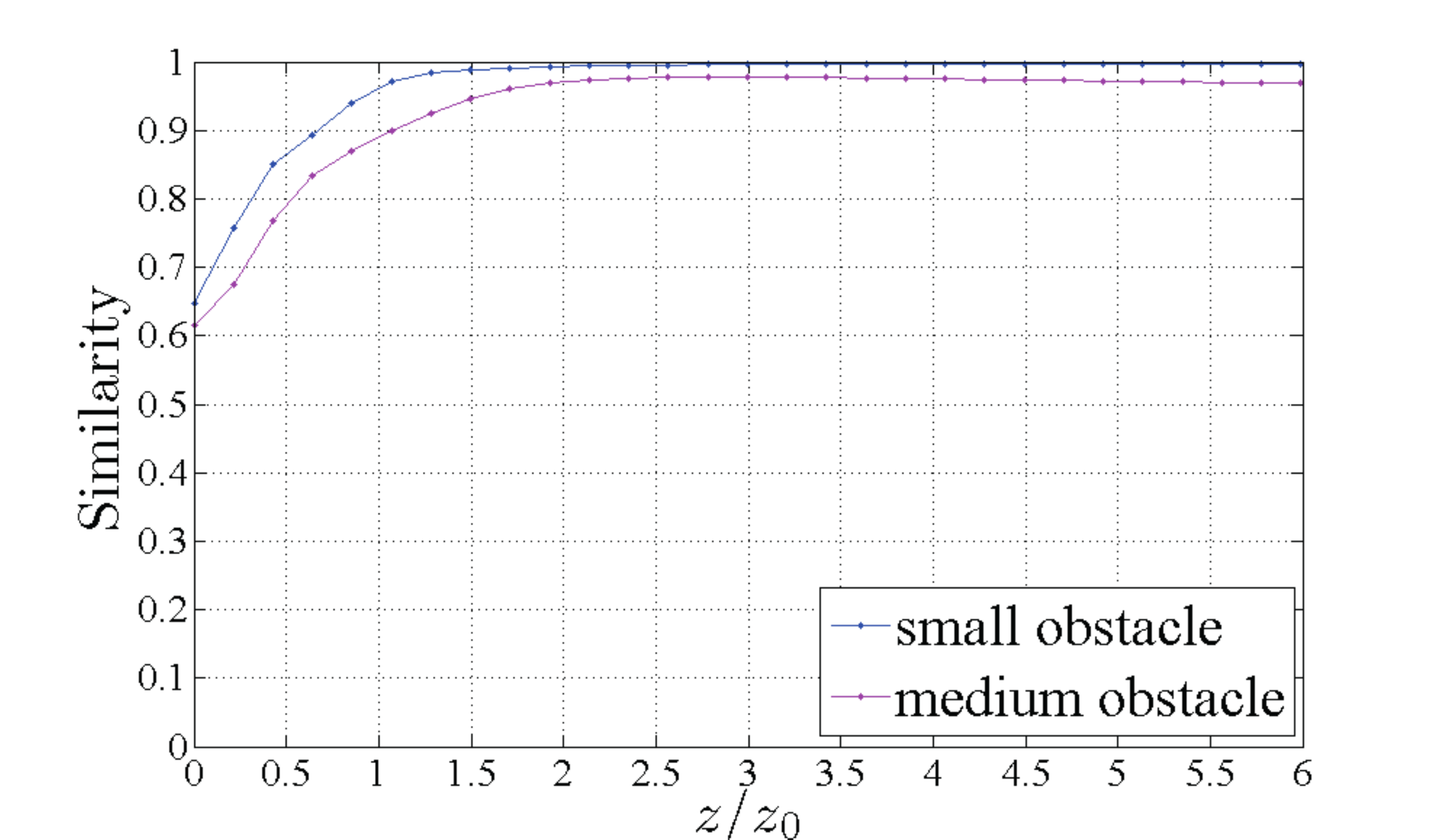}
  	\label{fig:simigb_ai}}
\caption{Behavior of RMS deviation (a,c) and Similarity function (b,d) within the internal domain $\varOmega$, at different propagation distances of  HG (a,b)  and IG (c,d) beams subject to different obstruction areas.}
\label{fig:rmssimai}
\end{figure}

To complement the results in Fig. \ref{fig:rmssimai}, we also computed the RMS and S for the already considered beams and obstructions employing the domain external to $\varOmega$ (within the area of $b^p(x,y)$).
It is interesting to note that the results, displayed in Fig. \ref{fig:rmssimae}, show a clear degradation of the external propagated fields. Fortunately, this degradation also saturates at certain propagation distances.
\begin{figure}[h!]
\centering
	\subfigure[]{
  	\includegraphics[width=6.4cm]{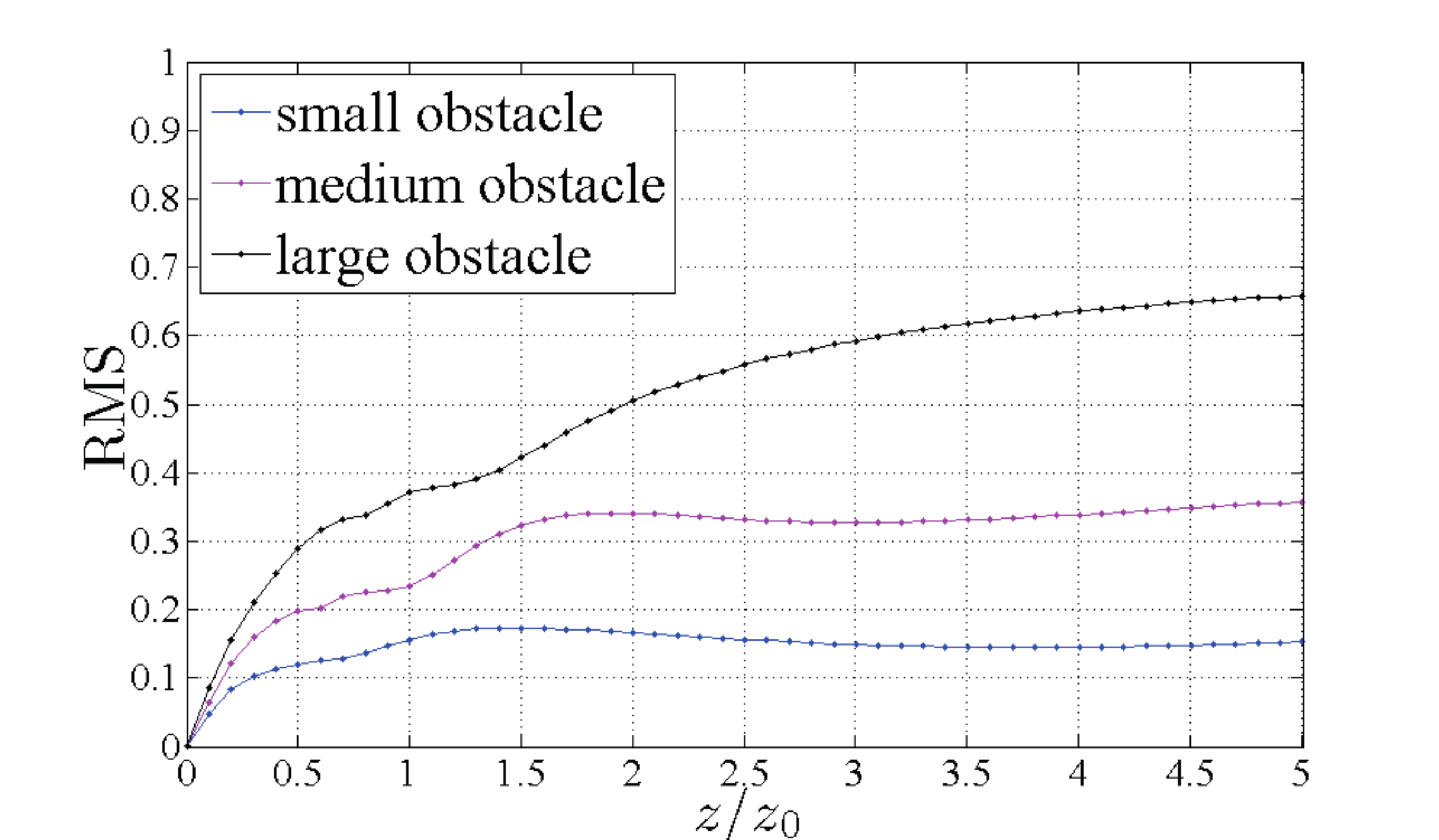}
  	\label{fig:rmshgb_ae}}
    \subfigure[]{
  	\includegraphics[width=6.4cm]{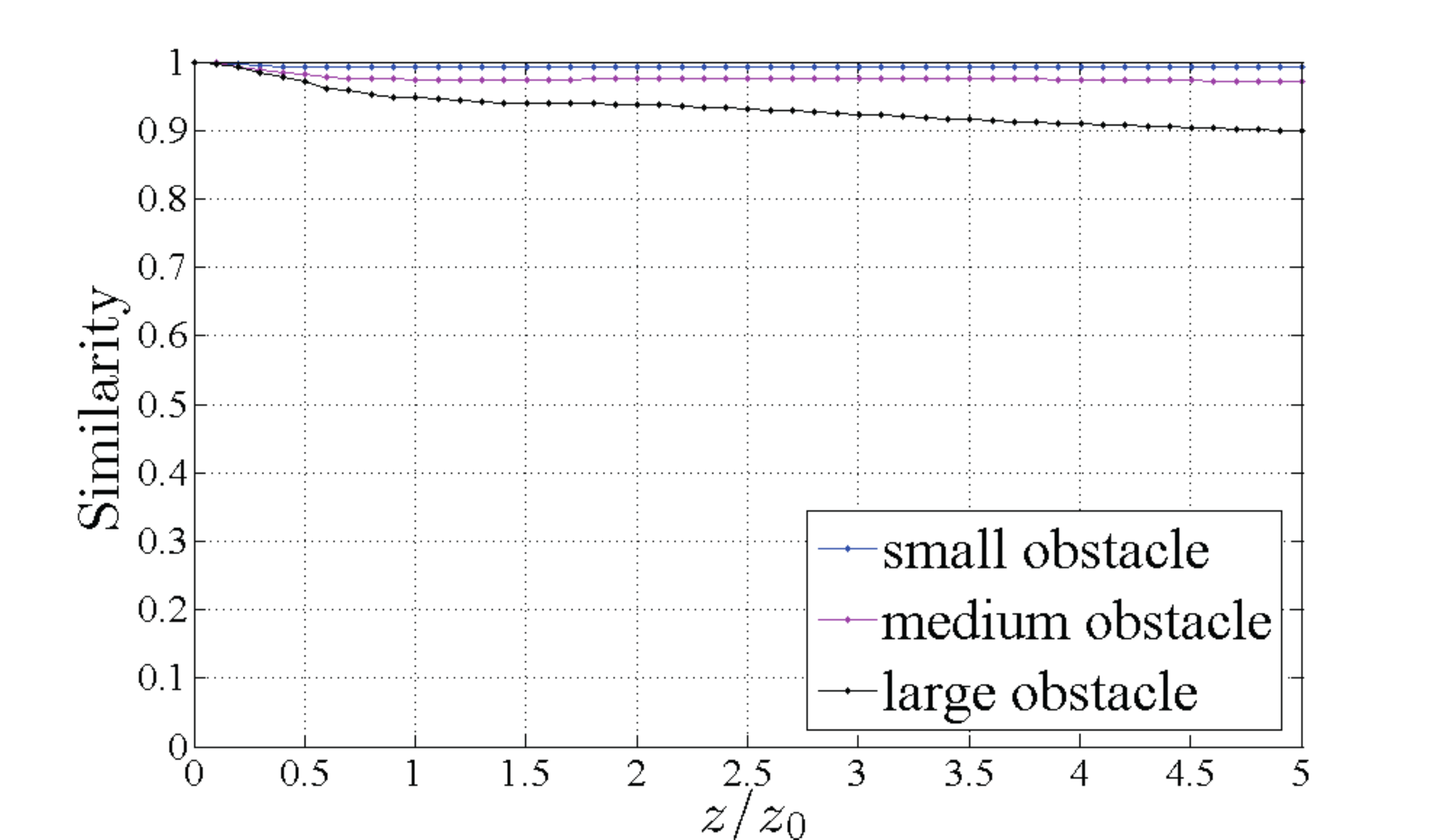}
  	\label{fig:simihgb_ae}}
 	\subfigure[]{
 	\includegraphics[width=6.4cm]{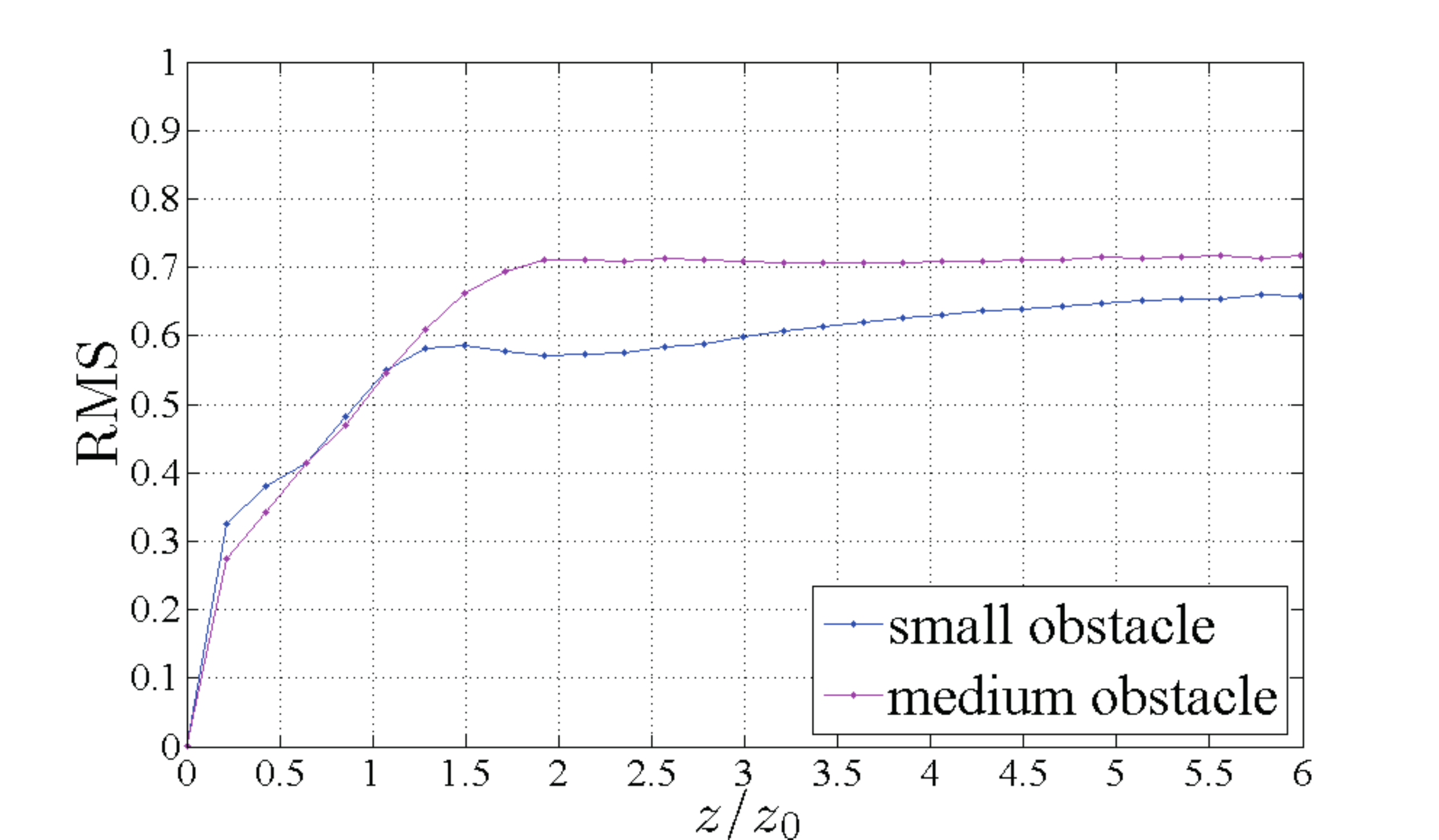}
   	\label{fig:rmsigb_ae}}
 	\subfigure[]{
  	\includegraphics[width=6.4cm]{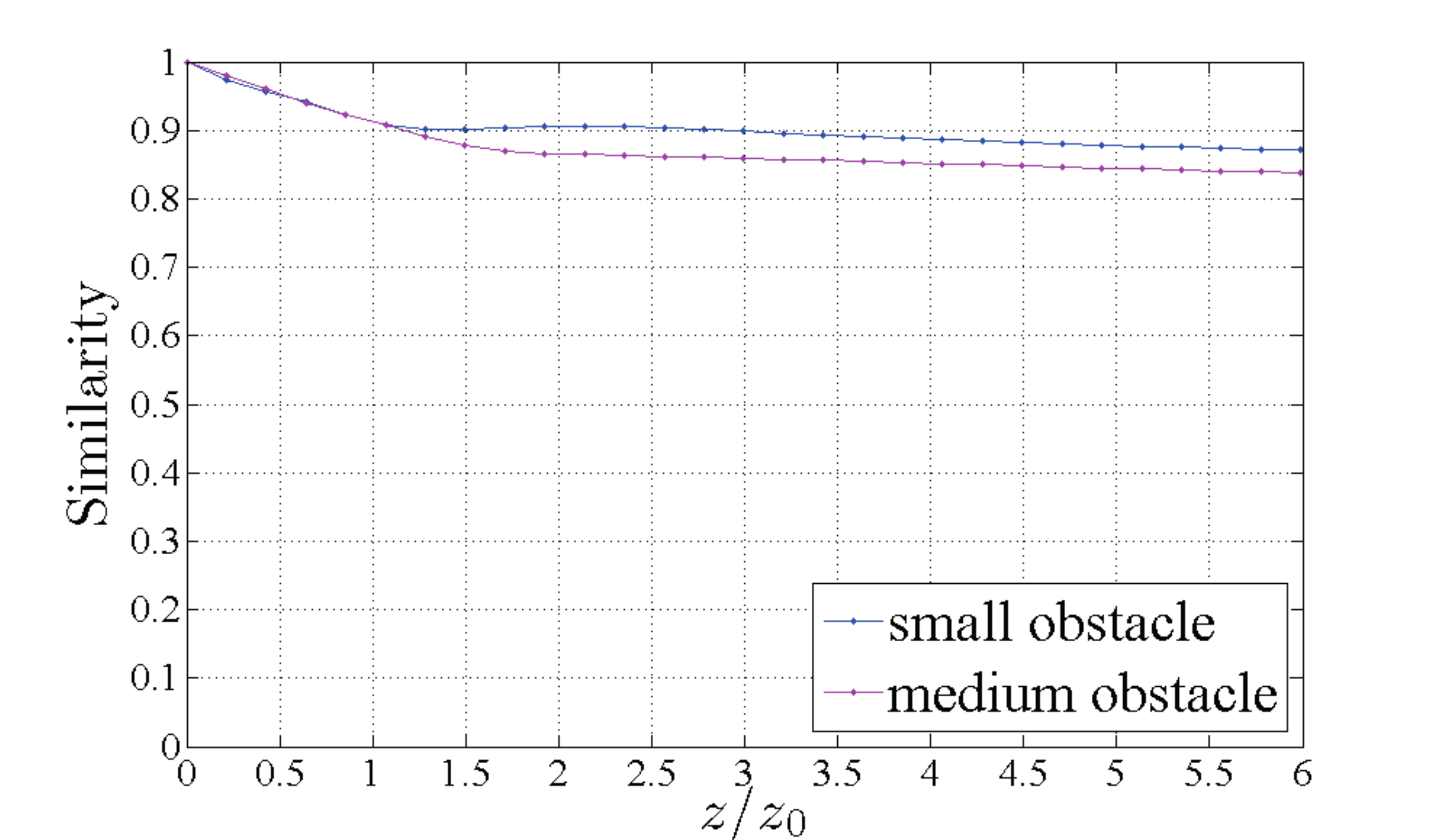}
  	\label{fig:simigb_ae}}
\caption{Behavior of RMS deviation (a,c) and Similarity function (b,d) outside the domain $\varOmega$ at different propagation distances of  HG (a,b)  and IG (c,d) beams subject to different obstruction areas.}
\label{fig:rmssimae}
\end{figure}
\section{Final remarks and conclusions}
We have analyzed the SH effect in SPIBs, a kind of propagation invariant beams with scale change. In particular we demonstrated numerically and experimentally the effect in the cases of HG and IG beams.

We established a semi-analytic parameter to asses the degree of SH, corresponding to the signal to noise intensity ratio ($Q$). This quantity is computed in terms of parameters that represent the features of both the beam and the obstruction applied to it. We obtained that for a given set of these parameters, there exist a restricted range of values for $Q$. Moreover, we presented a formula to compute the required propagation distance for each $Q$ value in the allowed range. This analysis establishes indirectly that the degree of SH in SPIBs can not be achieved beyond certain limit, regardless the propagation distance of the obstructed beam. Such restriction is confirmed qualitatively in the images of self-reconstructed beams, obtained numerically and experimentally, in section section 3 (figures \ref{fig:NH88om} to \ref{fig:Ni823op}).

To complement the semi-analytic assessment of SH by means of the parameter $Q$, we evaluated numerically the effect by computing  the RMS deviation and the Similarity for the obstructed propagated beams, respect to the non-disturbed beams. We first computed such parameters in a domain $\varOmega$ (referred as internal domain), which corresponds to a scaled version of the applied obstruction. The RMS deviations computed in this domain, present a reduction during propagation, up to a limit that is dependent on the beam and obstruction features. The appearance of this limit roughly provides a quantitative confirmation of the limit in the SH, that was analytically established by means of the $Q$ ratio. Although the similarity $S$ is less sensible that the RMS, the SH limit can also be noted at close views of the $S$ plots (which for brevity are not shown).

We also computed the RMS and $S$ in a domain external to $\varOmega$ (within the area of the beam $b^p(x,y)$). The results (in Fig. \ref{fig:rmssimae}) are interesting, although not surprising. They show that  the RMS increases and the similarity reduces at the the external domains.

The analytical discussion regarding the signal to noise intensity ratio ($Q$) is an original contribution in  this paper that can  be extended to analyze the SH in other propagation invariant beams (Bessel, Airy, etc). Another contribution has been the separate evaluation of the RMS and similarity in an internal and an external domain. The results in this context improves the understanding of the SH effect.
\end{document}